\documentclass[aps,prl,twocolumn,superscriptaddress,showpacs]{revtex4}
\usepackage{epsfig}
\usepackage{subfigure}
\newcommand{\sgeq}   {\raisebox{-.6ex}{${\textstyle\stackrel{>}{\sim}}$}}
\newcommand{\sleq}   {\raisebox{-.6ex}{${\textstyle\stackrel{<}{\sim}}$}}

\newcommand{\W}{{\it W}}
\newcommand{\Z}{{\it Z}}
\newcommand{\ee}     {\mbox{${e}^+{e}^-$}}
\newcommand{\mumu}   {\mbox{$\mu^+ \mu^-$}}

\newcommand{\pppZp}     {\mbox{$p\bar{p}\rightarrow{p}\Z\bar{p}$}}
\newcommand{\ppnWp}     {\mbox{$p\bar{p}\rightarrow{n}\W\bar{p}$}}
\newcommand{\ggll}  {$\gamma\gamma\rightarrow \ell^+\ell^-$}
\newcommand{\ggmm}  {$\gamma\gamma\rightarrow \mu^+\mu^-$}
\newcommand{\ggee}  {$\gamma\gamma\rightarrow e^+e^-$}

\newcommand{\pppggpll}   {p\bar{p}\rightarrow p \gamma\gamma\bar{p} \rightarrow p\ell^+\ell^-\bar{p}}
\newcommand{\qq}     {$q\bar{q}$}
\newcommand{\pp}     {$p\bar{p}$}
\newcommand{\ellell} {\mbox{$\ell^+ \ell^-$}}
\newcommand{\Zll}  {\mbox{${Z}\rightarrow \ell^+ \ell^-$}}
\newcommand{\Zgll}  {\mbox{${Z/\gamma^{*}}\rightarrow \ell^+ \ell^-$}}
\newcommand{\Wlnu}     {\mbox{${W}\rightarrow{\ell}\nu$}}

\newcommand{\yZ}    {$y_{Z}$}
\newcommand{\ptll}{p_{\rm T}(\ell\ell)}

\newcommand{\Mll}  {M_{\ell\ell}}

\newcommand{\crossexcl} {\sigma_{excl}(Z)}
\newcommand{\dsdy} {\frac{d\sigma}{dy}|_{y=0}}

\newcommand{\GeVcsq} {\mbox{${\rm GeV}/c^{2}$}}
\newcommand{\GeVc} {\mbox{${\rm GeV}/c$}}

\newcommand{\lpair} {\textsc{lpair}}
\newcommand{\pythia} {\textsc{pythia}}
\newcommand{\photos} {\textsc{photos}}
\newcommand{\geant} {\textsc{geant}}

\newcommand{\intLeff}    {\mbox{$\int \mathcal{L}_{eff}$}}

\newcommand{\pt}{$p_T$}
\newcommand{\ptl}{$p_T^{\ell}$}
\newcommand{\dphi}{$\Delta\phi_{\ell\ell}$}

\newcommand{\sume}{$\Sigma{E}$}
\begin{document}
\lefthyphenmin=2
\righthyphenmin=3

%new 060600
%\setlength{\baselineskip}{24pt}
%\preprint{FERMILAB-Pub-97/109-E}
%
% ======> Title of the paper goes here <====================
%
\title{
Search for exclusive \Z\ boson production and observation of high mass
$\pppggpll$ events in $p\bar{p}$ collisions
at $\sqrt{s}$ = 1.96 TeV
}
\affiliation{Institute of Physics, Academia Sinica, Taipei, Taiwan 11529, Republic of China} 
\affiliation{Argonne National Laboratory, Argonne, Illinois 60439} 
\affiliation{University of Athens, 157 71 Athens, Greece} 
\affiliation{Institut de Fisica d'Altes Energies, Universitat Autonoma de Barcelona, E-08193, Bellaterra (Barcelona), Spain} 
\affiliation{Baylor University, Waco, Texas  76798} 
\affiliation{Istituto Nazionale di Fisica Nucleare Bologna, $^v$University of Bologna, I-40127 Bologna, Italy} 
\affiliation{Brandeis University, Waltham, Massachusetts 02254} 
\affiliation{University of California, Davis, Davis, California  95616} 
\affiliation{University of California, Los Angeles, Los Angeles, California  90024} 
\affiliation{University of California, San Diego, La Jolla, California  92093} 
\affiliation{University of California, Santa Barbara, Santa Barbara, California 93106} 
\affiliation{Instituto de Fisica de Cantabria, CSIC-University of Cantabria, 39005 Santander, Spain} 
\affiliation{Carnegie Mellon University, Pittsburgh, PA  15213} 
\affiliation{Enrico Fermi Institute, University of Chicago, Chicago, Illinois 60637}
\affiliation{Comenius University, 842 48 Bratislava, Slovakia; Institute of Experimental Physics, 040 01 Kosice, Slovakia} 
\affiliation{Joint Institute for Nuclear Research, RU-141980 Dubna, Russia} 
\affiliation{Duke University, Durham, North Carolina  27708} 
\affiliation{Fermi National Accelerator Laboratory, Batavia, Illinois 60510} 
\affiliation{University of Florida, Gainesville, Florida  32611} 
\affiliation{Laboratori Nazionali di Frascati, Istituto Nazionale di Fisica Nucleare, I-00044 Frascati, Italy} 
\affiliation{University of Geneva, CH-1211 Geneva 4, Switzerland} 
\affiliation{Glasgow University, Glasgow G12 8QQ, United Kingdom} 
\affiliation{Harvard University, Cambridge, Massachusetts 02138} 
\affiliation{Division of High Energy Physics, Department of Physics, University of Helsinki and Helsinki Institute of Physics, FIN-00014, Helsinki, Finland} 
\affiliation{University of Illinois, Urbana, Illinois 61801} 
\affiliation{The Johns Hopkins University, Baltimore, Maryland 21218} 
\affiliation{Institut f\"{u}r Experimentelle Kernphysik, Universit\"{a}t Karlsruhe, 76128 Karlsruhe, Germany} 
\affiliation{Center for High Energy Physics: Kyungpook National University, Daegu 702-701, Korea; Seoul National University, Seoul 151-742, Korea; Sungkyunkwan University, Suwon 440-746, Korea; Korea Institute of Science and Technology Information, Daejeon, 305-806, Korea; Chonnam National University, Gwangju, 500-757, Korea} 
\affiliation{Ernest Orlando Lawrence Berkeley National Laboratory, Berkeley, California 94720} 
\affiliation{University of Liverpool, Liverpool L69 7ZE, United Kingdom} 
\affiliation{University College London, London WC1E 6BT, United Kingdom} 
\affiliation{Centro de Investigaciones Energeticas Medioambientales y Tecnologicas, E-28040 Madrid, Spain} 
\affiliation{Massachusetts Institute of Technology, Cambridge, Massachusetts  02139} 
\affiliation{Institute of Particle Physics: University of Alberta, Edmonton, Canada, T6G 2G7; McGill University, Montr\'{e}al, Qu\'{e}bec, Canada H3A~2T8; Simon Fraser University, Burnaby, British Columbia, Canada V5A~1S6; University of Toronto, Toronto, Ontario, Canada M5S~1A7; and TRIUMF, Vancouver, British Columbia, Canada V6T~2A3} 
\affiliation{University of Michigan, Ann Arbor, Michigan 48109} 
\affiliation{Michigan State University, East Lansing, Michigan  48824}
\affiliation{Institution for Theoretical and Experimental Physics, ITEP, Moscow 117259, Russia} 
\affiliation{University of New Mexico, Albuquerque, New Mexico 87131} 
\affiliation{Northwestern University, Evanston, Illinois  60208} 
\affiliation{The Ohio State University, Columbus, Ohio  43210} 
\affiliation{Okayama University, Okayama 700-8530, Japan} 
\affiliation{Osaka City University, Osaka 588, Japan} 
\affiliation{University of Oxford, Oxford OX1 3RH, United Kingdom} 
\affiliation{Istituto Nazionale di Fisica Nucleare, Sezione di Padova-Trento, $^w$University of Padova, I-35131 Padova, Italy} 
\affiliation{LPNHE, Universite Pierre et Marie Curie/IN2P3-CNRS, UMR7585, Paris, F-75252 France} 
\affiliation{University of Pennsylvania, Philadelphia, Pennsylvania 19104}
\affiliation{Istituto Nazionale di Fisica Nucleare Pisa, $^x$University of Pisa, $^y$University of Siena and $^z$Scuola Normale Superiore, I-56127 Pisa, Italy} 
\affiliation{University of Pittsburgh, Pittsburgh, Pennsylvania 15260} 
\affiliation{Purdue University, West Lafayette, Indiana 47907} 
\affiliation{University of Rochester, Rochester, New York 14627} 
\affiliation{The Rockefeller University, New York, New York 10021} 
\affiliation{Istituto Nazionale di Fisica Nucleare, Sezione di Roma 1, $^{aa}$Sapienza Universit\`{a} di Roma, I-00185 Roma, Italy} 
\affiliation{Rutgers University, Piscataway, New Jersey 08855} 
\affiliation{Texas A\&M University, College Station, Texas 77843} 
\affiliation{Istituto Nazionale di Fisica Nucleare Trieste/Udine, I-34100 Trieste, $^{bb}$University of Trieste/Udine, I-33100 Udine, Italy} 
\affiliation{University of Tsukuba, Tsukuba, Ibaraki 305, Japan} 
\affiliation{Tufts University, Medford, Massachusetts 02155} 
\affiliation{Waseda University, Tokyo 169, Japan} 
\affiliation{Wayne State University, Detroit, Michigan  48201} 
\affiliation{University of Wisconsin, Madison, Wisconsin 53706} 
\affiliation{Yale University, New Haven, Connecticut 06520} 
\author{T.~Aaltonen}
\affiliation{Division of High Energy Physics, Department of Physics, University of Helsinki and Helsinki Institute of Physics, FIN-00014, Helsinki, Finland}
\author{J.~Adelman}
\affiliation{Enrico Fermi Institute, University of Chicago, Chicago, Illinois 60637}
\author{T.~Akimoto}
\affiliation{University of Tsukuba, Tsukuba, Ibaraki 305, Japan}
\author{M.G.~Albrow}
\affiliation{Fermi National Accelerator Laboratory, Batavia, Illinois 60510}
\author{B.~\'{A}lvarez~Gonz\'{a}lez$^q$}
\affiliation{Instituto de Fisica de Cantabria, CSIC-University of Cantabria, 39005 Santander, Spain}
\author{S.~Amerio$^w$}
\affiliation{Istituto Nazionale di Fisica Nucleare, Sezione di Padova-Trento, $^w$University of Padova, I-35131 Padova, Italy} 
\author{D.~Amidei}
\affiliation{University of Michigan, Ann Arbor, Michigan 48109}
\author{A.~Anastassov}
\affiliation{Northwestern University, Evanston, Illinois  60208}
\author{A.~Annovi}
\affiliation{Laboratori Nazionali di Frascati, Istituto Nazionale di Fisica Nucleare, I-00044 Frascati, Italy}
\author{J.~Antos}
\affiliation{Comenius University, 842 48 Bratislava, Slovakia; Institute of Experimental Physics, 040 01 Kosice, Slovakia}
% ...

\collaboration{CDF Collaboration}
\noaffiliation

%%%%%% end \input{mumuauthors.tex}

\begin{abstract}
We present a search for exclusive \Z\ boson production in proton-antiproton 
collisions at $\sqrt{s}$ = 1.96 TeV, using the CDF~II detector at Fermilab. 
We observe no exclusive \Zll\ candidates and place the first upper
limit on the exclusive \Z\ cross section in hadron collisions, 
$\crossexcl <$ 0.96~pb at 95\% confidence level.
In addition, we observe eight candidate exclusive dilepton events from
the quantum electrodynamic process $\pppggpll$, and measure
the cross section for $\Mll >$~40~\GeVcsq\ and
$|\eta_{\ell}| < 4$ to be $\sigma = 0.24^{+0.13}_{-0.10}$~pb, which is
the first measurement for this mass range and is
consistent with the standard model prediction.
%%%%% end \input{prlz_abs.tex}
\end{abstract} 

\maketitle

\begin{figure}[htbp]
\centering
  \subfigure[]{  
    \epsfig{file=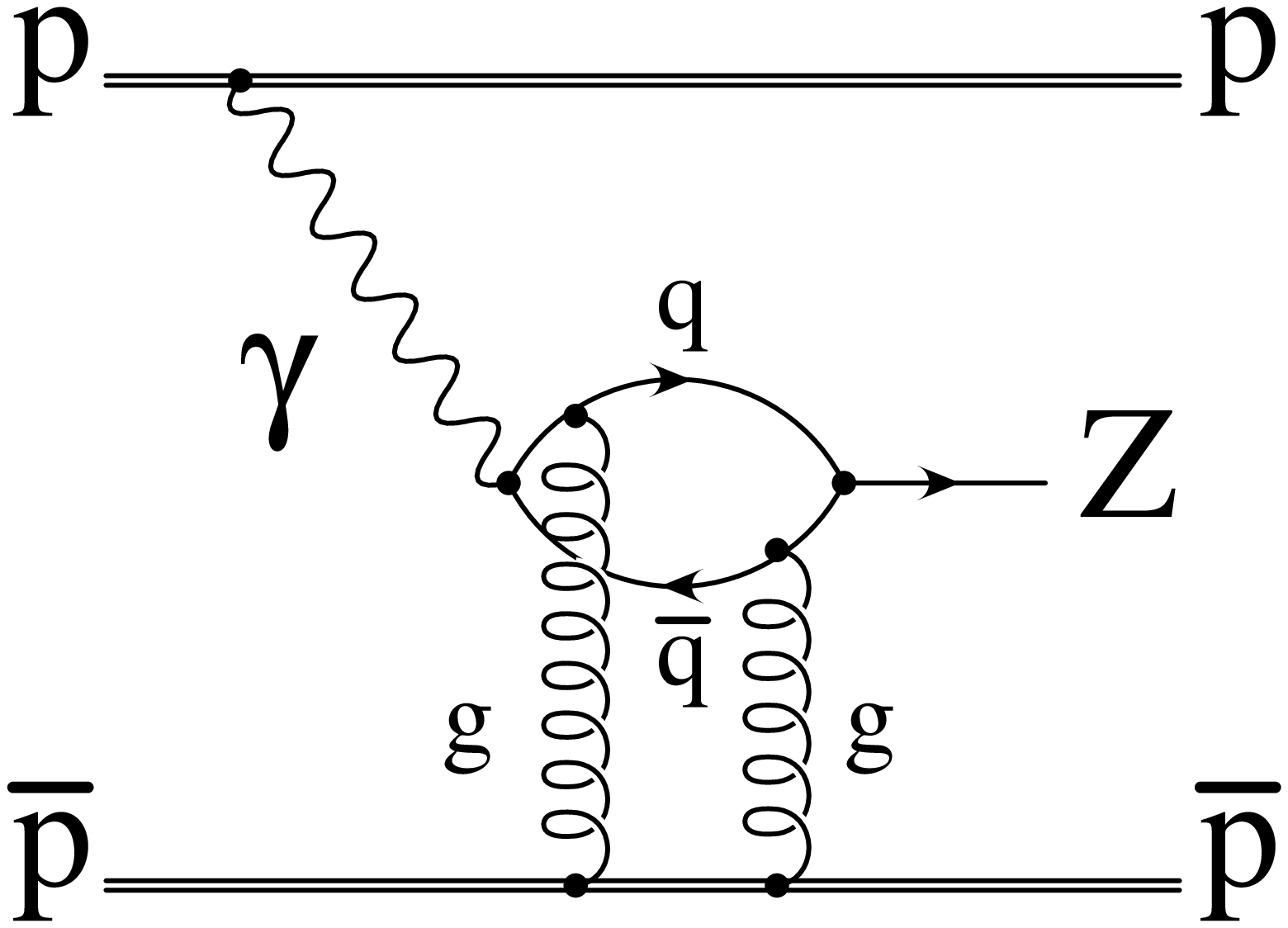,width=4cm}
    \label{fig:feyn-Z}
   }
  \subfigure[]{  
    \epsfig{file=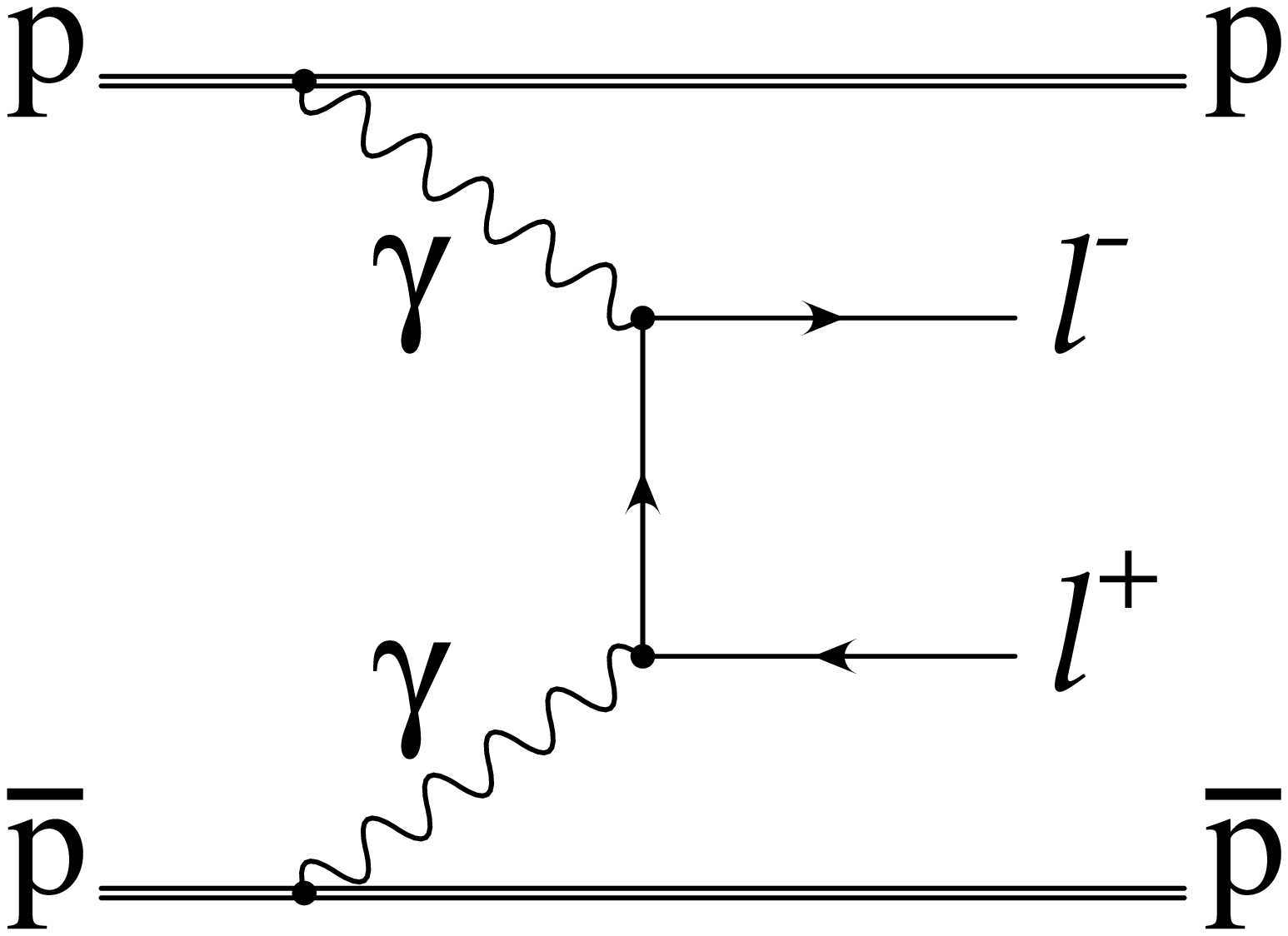,width=4cm}
    \label{fig:feyn-qed}
   }
  \caption{(a) Exclusive photoproduction of a \Z\ boson and (b) exclusive dilepton production via two-photon exchange at the Tevatron.}
  \label{fig:feyn}
\end{figure}

At the Tevatron \pp\ collider it is possible to produce \Z\ bosons exclusively, in association with no other particles
except the $p$ and $\bar{p}$: \pppZp.
%and is much rarer than inclusive \Z\ production, where the hadrons break up, and single diffractive \Z\ production.
The colliding hadrons emerge intact with small transverse momenta, \pt~\cite{ref:kine}.
% and the \Z\ is emitted at small rapidity, \yZ. 
The process is predicted by the standard model (SM) to proceed via photoproduction. 
A radiated virtual photon fluctuates to a \qq\ loop which scatters elastically by two-gluon exchange on the
 (anti)proton and materializes as a \Z, as shown in Figure~\ref{fig:feyn-Z}.
%
%\begin{figure}[htbp]
%  \begin{center}
%    \includegraphics[width=0.5\textwidth]{feynmanp.ps}
%    \caption{Feynman diagrams for (a) $\gamma\gamma\rightarrow l^+l^-$, (b) $\gamma$\pom$\rightarrow Z$.
%    EDIT LEFT 2, REMOVE 3rd.
% \label{fig:emet} }
%  \end{center}
%\end{figure}hi
The same mechanism gives photoproduction of the vector mesons $J/\psi, \psi(2S)$ and $\Upsilon$, which have been studied in $ep$ collisions
at HERA~\cite{ref:hera} and recently observed for the first time in $p\bar{p}$  collisions by CDF~\cite{ref:exclmm}. 
The SM cross section for exclusive \Z\ production is predicted to be 
$\sigma_{excl}(Z) = $~0.3~fb~\cite{ref:motyka},
% (not including branching fractions to lepton pairs, 3.36\% per flavor) 
and is thus below the threshold for detection.
%too small to be currently detected.
%At the large hadron collider (LHC) the cross section is predicted to be much larger, 13 fb by Ref.~\cite{ref:motyka} and 69 fb by Ref.~\cite{ref:goncalves}.
% and perhaps detectable with forward proton detectors added to ATLAS or CMS~\cite{fp420}. 
% DO WE NEED THE FORWARD PROTON TAGGERS TO DETECT THIS?
An observation at the Tevatron would therefore be evidence for
beyond SM (BSM) physics. 
A BSM theory of the pomeron $I\!\!P$~\cite{ref:white}
predicts a much larger cross section, possibly orders of magnitude larger, but without a quantitative estimate. 
In this theory the pomeron 
%is mostly a reggeized gluon and a sea of wee gluons, and it 
couples strongly to the electroweak sector through a pair
of color sextet quarks which contribute to the quark loop shown in Figure~\ref{fig:feyn-Z}.
%At the large hadron collider (LHC) the SM cross section is predicted to be 13 fb~\cite{ref:motyka}.
%Again, a much larger cross section is predicted  by Ref.~\cite{ref:white}.
%In addition to an enhanced $\gamma I\!\!P Z$ coupling, one expects a large cross section for
%$I\!\!PI\!\!P \rightarrow W^+W^-$ (or $ZZ$)  at the LHC.

This Letter presents a search for exclusive \Z\ production with the \Z\ decaying to a \mumu\ or \ee\ pair, 
and a measurement of the  cross section for exclusive \mumu\ and \ee\ production with dilepton invariant mass $\Mll >$ 40
\GeVcsq\ and $|\eta_{\ell}| < 4$. We use the CDF II detector at the Tevatron with \pp\ collisions at a center of mass energy $\sqrt{s} = 1.96$~TeV.
The exclusive dilepton process is expected in quantum electrodynamics (QED) through $\pppggpll$, as shown in Figure~\ref{fig:feyn-qed}.
For the remainder of this Letter we will refer to this process as \ggll\ for convenience.
We have previously observed \ggee\ with $10 < M_{ee}< 40$~\GeVcsq~\cite{ref:exclee} and
\ggmm\ with $3 < M_{\mu\mu}< 4$~\GeVcsq~\cite{ref:exclmm} and measured cross sections
in agreement with expectations.  
The final state particles in exclusive dilepton events are identical to those in exclusive \Z\ production with leptonic decay, the only difference in the signature being the $\Mll$ distribution and other kinematics.
Agreement with the precise theoretical prediction therefore gives us confidence in our sensitivity to selecting exclusive \Z\ bosons.
%At the LHC, high mass \ggll\ events could provide a powerful tool for 
%calibrating forward proton spectrometers, as both proton momenta are well %known from the dilepton kinematics. The cross section is also
%known to $<$ 0.3\% and therefore the process may provide a calibration of the %machine luminosity. 

CDF II is a general purpose detector which is described in detail elsewhere~\cite{ref:cdf}; 
here we give a brief summary. 
Surrounding the collision region is a tracking system consisting of
silicon microstrips and a cylindrical drift
chamber, the central outer tracker (COT), in a 1.4~Tesla solenoid.
The tracking system
tracks particles with  $p_T$~\sgeq~0.3~GeV/c and pseudorapidity $|\eta|$~\sleq~2~\cite{ref:kine}.
Central and end-plug calorimeters cover the
range $|\eta|<1.3$ and $1.3<|\eta|<3.6$ respectively, with separate electromagnetic (EM) and hadronic (HAD) compartments. 
Outside the calorimeters, drift
chambers measure muons in the region $|\eta| <$ 1.0.
The regions $3.6<|\eta|<5.2$ are covered by
lead-liquid scintillator calorimeters called the
miniplugs~\cite{ref:miniplug}. At higher pseudorapidities,
$5.4<|\eta|<7.4$, scintillation counters called beam shower counters (BSC) are located along the beam pipe.  Gas
$\check{\textrm{C}}$erenkov detectors covering 3.7 $<|\eta|<$ 4.7 measure the luminosity with a 6\%
uncertainty by counting inelastic interactions~\cite{ref:clc}.
Tracking detectors in a Roman pot spectrometer~\cite{ref:rp}, located 57 meters from the interaction point, can detect antiprotons with small $p_T$ and
 0.03~$\sleq~\xi(\bar{p})~\sleq$~0.08, where $\xi(\bar{p})$ is the fractional momentum loss of the antiproton~\cite{ref:xi}. These detectors were operational for approximately 30\%\ of the data set used in this analysis.
%Tracking detectors in roman pots, present
%for part of the data taking and with coverage 0.03$\sleq \xi(\bar{p}) \sleq$ %0.08, 
%can detect some scattered antiprotons; here $\xi(\bar{p})$ is the fractional energy loss of the antiproton~\cite{ref:kine}.
%$1 - \frac{p_z}{p_{beam}}$ where 
%for part of the data taking and with coverage 0.03$\lesssim\xi(\bar{p})\lesssim$ 0.08~\cite{ref:kine} can detect some scattered antiprotons, 
%There are no detectors able to measure forward protons.

For the \ggll\ event selection we select a sample of \ellell\ pairs in a kinematic region where this process has not previously been observed, with $\Mll > 40$~\GeVcsq\ and lepton transverse momenta
\ptl\ $>$ 20~\GeVc. For the exclusive \Z\ search we select a subsample with an invariant mass close to the \Z\ mass, 82 $< \Mll <$ 98~\GeVcsq, and \ptl\ $>$ 25~\GeVc.
The \mumu\ events are collected with a trigger requiring one muon with \pt\ $>$ 18 \GeVc.
Offline we require two candidate muons. One muon must be detected in the COT, the central calorimeter, and
the muon chambers, and therefore has $|\eta_{\mu}|<1.0$. To increase acceptance the second muon is only required to be detected in the COT and therefore has $|\eta_{\mu}|<1.5$.
Events consistent with cosmic rays are eliminated with an identification algorithm~\cite{ref:cosmic} that uses the timing of the COT drift chamber hits.
The muon kinematics are found from the COT track momentum measurement.
The \ee\ events are collected with a trigger requiring one central electron with \pt\ $>$ 18 \GeVc.
Offline we require one candidate electron reconstructed in the central EM calorimeter and matched to a COT track,
and a second electron reconstructed either in the same way or in the end-plug EM calorimeter where no matching COT track is required, since
the tracking efficiency is lower in this region. The central electrons have  $|\eta_{e}| < 1.3$ and the end-plug electrons
have $1.3 < |\eta_{e}| < 3.6$.
The electron kinematics are found from the calorimeter energy measurement, but if a track is matched to the calorimeter cluster it is used 
to determine the electron direction. If no track is matched the $z$ position of the interaction is measured from the other electron track, and is used to determine the kinematics. 
Events with two central electrons are denoted CC \ee\ events, and those with one central and one end-plug electron
are denoted CP \ee\ events.  The \mumu, CC \ee, and CP \ee\ events are each treated as separate final states which are ultimately
combined together.
With an integrated luminosity of  2.20(2.03) fb$^{-1}$ in the electron(muon) channels we find a total of 317,712 candidate dileptons with 
$\Mll >$ 40~\GeVcsq, of which 183,332 are in the \Z\ region  82 $< \Mll <$ 98~\GeVcsq.

Starting with the dilepton samples we select events that are consistent with arising from exclusive production, by requiring
 that no other particles are produced in the collision.
We veto events where any additional tracks are reconstructed in the COT or the silicon tracker, or 
where any of the calorimeters have a total energy deposition above that expected from noise. 
For this purpose the calorimeters are divided into five sub-detectors and the energy of all towers is summed,
excluding those traversed by and surrounding the charged leptons, to give five \sume\ values.
Each \sume\ is required to be less than a threshold, which is
determined by studying two control samples: 
(1) events selected with a random bunch-crossing (zero bias) trigger with no tracks in the event, 
which should give distributions dominated by noise and 
(2) \Wlnu\ events with no detected tracks other than one from the charged lepton, which should give the distributions
expected for non-exclusive \Zll\ events with no additional tracks. 
The production mechanism for non-exclusive \W\ bosons is very similar to that for \Z\ bosons and the 
cross section for exclusive \W\ production (\ppnWp) is negligible, making them an excellent control sample. 
The chosen energy cuts  are \sume~$<$~3~GeV in each of the East and West miniplugs, $<$~5~GeV in both EM plugs, $<$~7~GeV in both HAD plugs, and $<$~0.35 GeV in the central EM calorimeter. 
%accept only 0.00009\% of inclusive \W\ events. 

These exclusivity cuts reject exclusive events that are in coincidence with additional inelastic \pp\ collisions.
It is therefore necessary to define an \emph{effective integrated luminosity} \intLeff.
The fraction of bunch crossings,
selected from the zero bias trigger, that pass the exclusivity cuts is
used to establish that \intLeff\ is 20.6\% of the total integrated luminosity.
The efficiency is found from distributions reweighted to account for the difference in the instantaneous luminosity profiles
between the zero bias events and the \Z\ events.
As a cross check we also estimate \intLeff\ using Poisson statistics and the mean number of expected interactions
per bunch crossing as a function of instantaneous luminosity to be 18.7\% of the total integrated luminosity. 
%This method gives a fraction that is 9\%\ lower than the previous method.
Since the method using the zero bias data properly takes into account events with no interactions that fail the cuts due to 
noise  in the calorimeters and fake reconstructed tracks, and events with a very soft interaction that pass the exclusivity cuts, we use 20.6\%\ to determine $\intLeff$, and take
the 9\%\ difference between the two methods as a systematic uncertainty.
%using $\sigma_{inel}$ = 60.7 mb, the bunch crossing rate (1.67 MHz) and the distribution of bunch luminosity (18.7\%). 
We find \intLeff\ = 
%($403 \pm 38 \pm 24$)~pb$^{-1}$ and ($467 \pm 41\pm 28$)~pb$^{-1}$ 
($403 \pm 45$)~pb$^{-1}$ and ($467 \pm 50$)~pb$^{-1}$ 
for the \mumu\ and \ee\ samples, respectively, 
where the uncertainty includes contributions from 
the difference between the two methods and the 6\% uncertainty on the 
CDF luminosity measurement.

In order to reduce the background from \ggll\ events where the proton
dissociates into forward-going hadrons,
% undergoes Coulomb dissociation,
we also make cuts on hits in the BSC detectors.
%The thresholds are found by comparing the individual photomultiplier (PMT) signals for 
%empty zero bias events with zero bias events where an interaction occurred and where 
%particles were detected in the surrounding BSC PMT counters. 
An event is vetoed if any photomultiplier has hits above threshold, which occurs in 76\%\ of zero bias
events that pass all the other exclusivity cuts. 
This inefficiency is included in the acceptance.
%by studying the ADC count distribution of the PMT with the highest ADC count in each BSC sub-detector

A total of eight events pass the \ggll\ selection criteria and no events pass the tighter exclusive \Zll\ criteria.
We use these events to measure the cross section for the \ggll\ process and we set an upper limit on the cross section for
exclusive \Z\ production. To do this we need to determine the acceptance for reconstructing the events
and the expected number of background events.

We calculate the acceptance for reconstructing \ggll\ events using the \lpair~\cite{ref:lpair} Monte Carlo (MC) event generator
together with a \geant~\cite{ref:geant} simulation of the CDF detector. 
We apply corrections to account for changes in the acceptance due to 
%higher order QED effects giving rise to 
internal Bremsstrahlung from the leptons, using the \photos~\cite{ref:photos} MC event generator.
The acceptance for the exclusive \Z\ search is found from the \pythia~\cite{ref:pythia} MC event generator, which simulates non-exclusive \Zgll\ events. 
Corrections are applied to account for the difference in kinematics between non-exclusive and exclusive production. We consider the \Z\ \pt\ distribution, which is assumed to be between 0 and 2~\GeVc\ for exclusive \Z\ production, the  \Z\ rapidity \yZ\ distribution, obtained from Ref.~\cite{ref:motyka}, and the angular 
distribution of the leptons. 
The latter is assumed to be
that for the decay of a spin-1 boson into two spin-1/2 fermions, which has the form (1 + $\cos^{2}\theta^{*}$), where $\theta^{*}$ is the polar angle between the outgoing lepton and the proton direction in the boson rest frame.
%The 76\%\ efficiency of the BSC cuts is also included in the acceptance.

The backgrounds to  \ggll\ events are non-exclusive \Zgll\ events that pass the exclusivity cuts, and \ggll\ events where the proton or antiproton dissociates
and the products are not detected in the forward detectors.
The former is found to be 0.28~$\pm$~0.19 events by assuming the fraction of non-exclusive \Zgll\ events passing the exclusivity cuts to be the same as that for non-exclusive \Wlnu\ events.
This fraction is found from \Wlnu\ data samples, selected by requiring a high \pt\ lepton and large missing transverse energy, 
to be $(9 \pm 6)\times{10^{-7}}$, where the uncertainty comes from the statistics of the samples.
The latter is found from the \lpair\ event generator, which also simulates \ggll\ events where one or both (anti)proton dissociate. 
We use the minimum bias Rockefeller MC~\cite{ref:mbr}, which fragments the excited (anti)proton into a nucleon and pions, to predict
the fraction of dissociation events that fail our exclusivity cuts due to particles in the region $|\eta| < 7.4$, which is the edge of the 
BSC acceptance.
We predict a total background of 1.45 $\pm$ 0.61 events, where the uncertainty comes from varying the exclusivity cuts and observing how
the number of events changes.
%******
%If we remove the BSC hits the number of candidate events increases from eight to 12. 
%The number of additional  \ggll\ events where the proton or antiproton dissociates is predicted by \lpair\ to be 6.9, 
%while these numbers 
%******

The backgrounds to exclusive \Z\ events are non-exclusive \Zgll\ events that pass the exclusivity cuts and exclusive \ggll\ events that 
have $\Mll$ in the \Z\ mass window. 
The former is found to be 0.163 $\pm$ 0.099 events using the method described above, and the latter is found from the \lpair\ MC samples to  be 0.492 $\pm$ 0.061 events.
We do not include a dissociation background for the exclusive \Z\ search; instead we quote an upper limit on the 
cross section for a \Z\ produced with no other particles with $|\eta| < 7.4$.

There is no evidence of cosmic ray contamination after using the
tagger described above.
This is verified by inspecting distributions such as the timing of the COT drift chamber hits and the acolinearity of the
lepton tracks. None of the candidate events are consistent with cosmic rays in any of the distributions.

We calculate a cross section for each final state  (\mumu, CC \ee\ and CP \ee) using the formula $$\sigma  = \frac{N - N_{bck}}{\alpha \times \intLeff},$$
where $N$ is the number of candidate events, $N_{bck}$ is the expected number of background events, and $\alpha$ is the acceptance.
Assuming equal rates for the \mumu\ and \ee\ processes, a combined cross section is found by forming a joint likelihood for the three final states, which is the product
of the Poisson probabilities to observe $N$ events in each final state.
The joint likelihood is a function of the signal cross section, and all the systematic uncertainties are 
nuisance parameters. The likelihood is maximized as a function of the signal cross section, 
and the uncertainty is obtained 
by multiplying this likelihood by a prior that is flat for positive cross sections
and finding the shortest interval containing 68\%\ of the integral.
%The inclusion of the systematics is done by averaging the posterior over variations of all parameters.
We treat the background and luminosity systematics as correlated and the acceptance systematics as uncorrelated.
The combined cross section for one lepton flavor is found to be $\sigma(\pppggpll) = 0.24^{+0.13}_{-0.10}$~pb for  $\Mll\ > 40~\GeVcsq$  and  $|\eta_{\ell}|<4$, 
which is in good agreement with the \lpair\ prediction of 0.256~pb.
%Here $\ell^{+}\ell^{-}$ signifies the decay to one leptonic flavor.

\begin{figure}[htbp]
\centering
\subfigure[]{  
    \epsfig{file=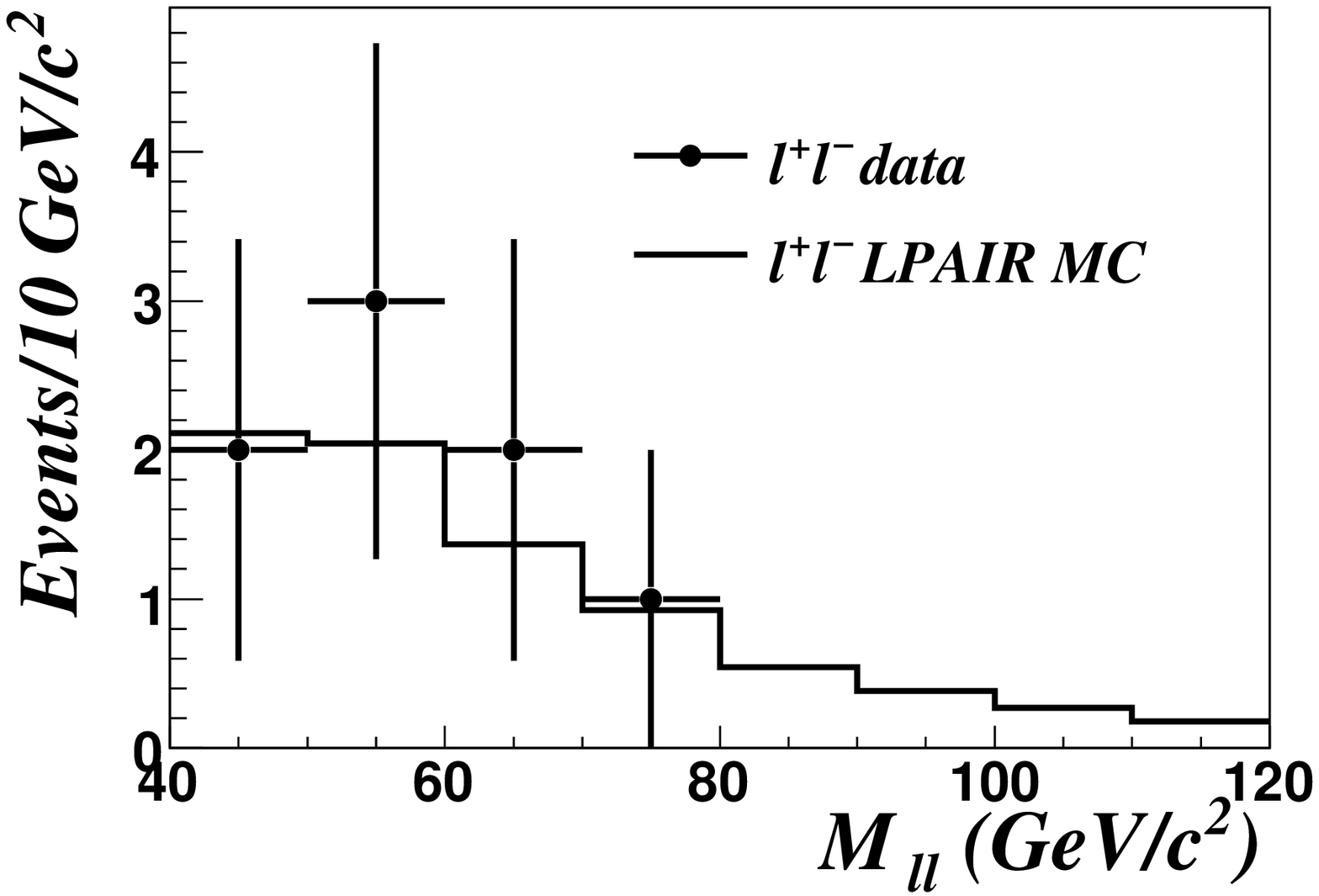,width=8cm}
    \label{fig:mass}
 }
\subfigure[]{  
 \epsfig{file=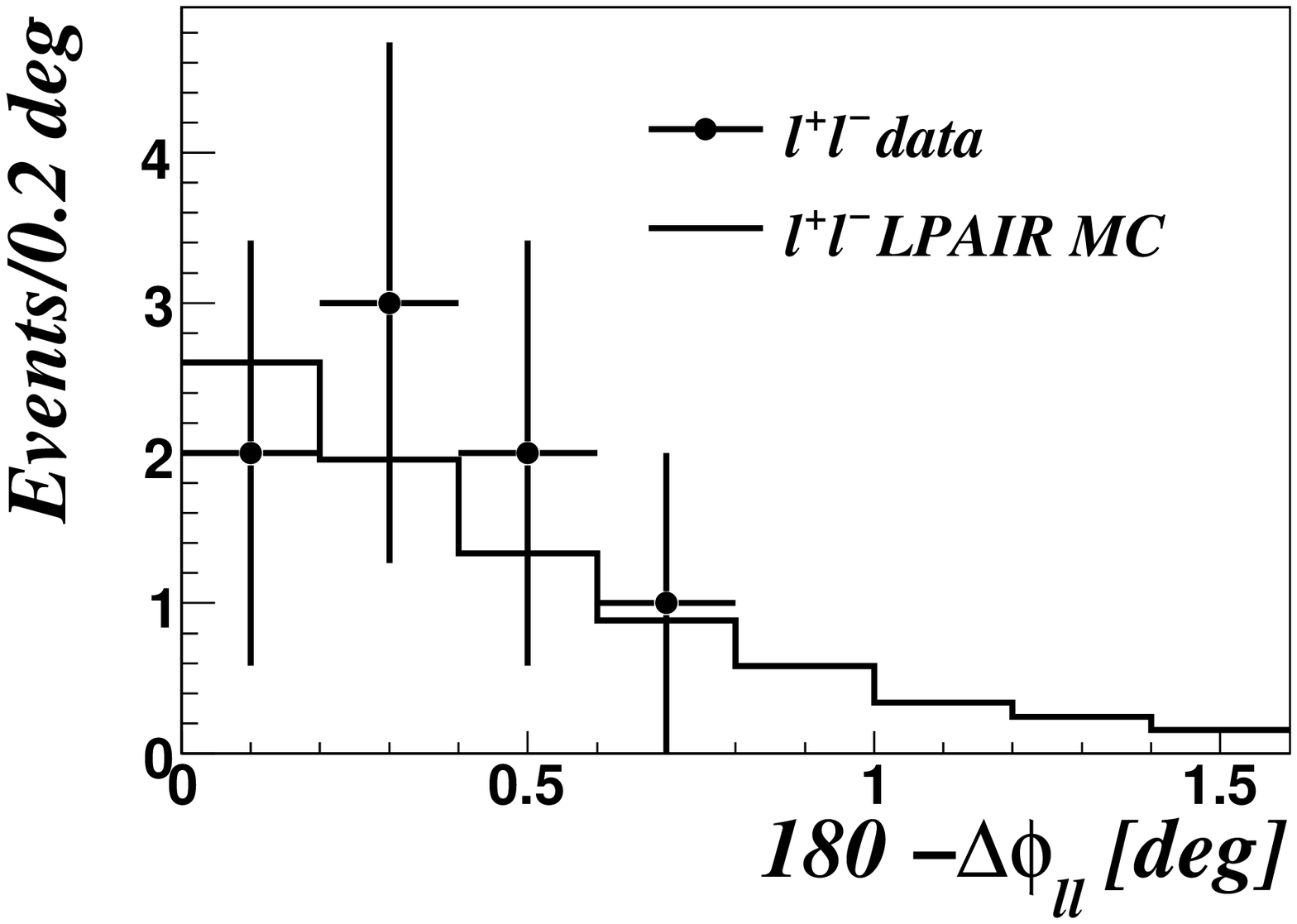,width=8cm}
 \label{fig:dphi}
 }
  \caption{(a) The dilepton invariant mass distribution, 
           and (b) the distribution of $180^{\circ}$ minus the difference in the azimuthal lepton angles
           for the data and the \lpair\ prediction with the \geant\ detector simulation, scaled to account
           for acceptance and luminosity.}
\end{figure}

Some of the kinematic properties of the candidate events are given in Table~\ref{tab:events},
where \ptl(1) and \ptl(2) are the lepton transverse momenta,
\dphi\ is the difference in the azimuthal lepton angles (i.e. 180$^{\circ}$ minus \dphi\ is the deviation from back-to-back in the transverse plane) 
and $\ptll$ is the \pt\ of the lepton pair. 
%The resolution on the lepton $\phi$ is approximately 0.01$^{\circ}$ and that on 
The resolution of the lepton transverse momenta is approximately 3.5(1.4)\%\ for electrons(muons).
All of the events have lepton pairs that are back-to-back in azimuth  with low $\ptll$ values, which is expected for \ggll\ events.
%The resolution in these quantities is estimated with a \textsc{geant}-based~\cite{geant} simulation of the CDF detector to be xx and xx respectively,
%with generator level (perfect resolution) values $\langle \Delta\phi \rangle$ = xx and 
%$\langle \ptll \rangle$ = xx GeV/c.   
\begin{table}[htbp]
\caption{Properties of the eight exclusive dilepton events, in order of $\Mll$.} 
\begin{center}
\begin{tabular}{cccccc}	
\hline \hline
Final state  & $\Mll$         & \ptl(1)  & \ptl(2)  &  $180^{\circ}$ - \dphi       & $\ptll$  \\
             & (\GeVcsq)      & (\GeVc)  & (\GeVc)  &  (degrees)   & (\GeVc) \\ 
\hline
\ee(CC)      & 40.9           & 20.4     & 20.1     & 0.38        & 0.26     \\ 
\ee(CC)      & 49.3           & 24.5     & 24.6     & 0.37        & 0.21     \\ 
\ee(CP)      & 50.4           & 20.5     & 20.2     & 0.05        & 0.31     \\ 
\ee(CP)      & 56.3           & 24.8     & 24.9     & 0.48        & 0.24     \\ 
\mumu\       & 58.6           & 24.1     & 24.4     & 0.17        & 0.32     \\ 
\mumu\       & 66.0           & 31.8     & 31.3     & 0.75        & 0.65      \\ 
\ee(CP)      & 67.1           & 24.1     & 24.0     & 0.51        & 0.24     \\ 
\ee(CP)      & 75.6           & 34.1     & 33.1     & 0.23        & 1.01     \\ 
\hline
\hline
\end{tabular}
\end{center}
\label{tab:events} 
\end{table}
Figures~\ref{fig:mass} and~\ref{fig:dphi} show the dilepton invariant mass and 180$^{\circ}$ minus \dphi\ distributions for the data together with the QED spectrum from \lpair\ and the
\geant\ detector simulation.
%scaled to account for the acceptance and luminosity of the combined data sets.
A good agreement with the data is observed.
%
%
%\begin{figure}[htbp]
%\centering
%    \epsfig{file=../plots/dphi_golden_tot_mc_prl.eps,width=9cm}
%  \caption{$\pi$ minus the difference in the azimuthal lepton angles (i.e. the deviation from back-to-back in the transverse plane) 
%           for the data and the \lpair\ prediction, which is scaled to account
%           for acceptance, luminosity and the \lpair\ cross section.}
%  \label{fig:dphi}
%\end{figure}
%

No events pass our exclusive \Zll\ selection criteria,
therefore we place an upper limit on the cross section of exclusive \Z\ production at the Tevatron.
We sum the three final states to give
$\sum{N}$ = 0, $\sum{N_{bck}}$ = 0.66~$\pm$~0.11,
and $\alpha\times\intLeff\times$BR(\ellell)  = 3.22~$\pm$~0.38 pb$^{-1}$.
Here we have used BR(\ellell) = 3.37\%\ as the branching fraction of the \Z\ to decay to one lepton flavor pair.
We use a Bayesian limit technique to set an upper  limit on the exclusive \Z\ cross section of $\crossexcl < 0.96$~pb at 95\%\ confidence level.
%We combine $\alpha$, $\intLeff$, and the 3.37\%\ branching fraction of the \Z\ to decay to a lepton pair
%to give $\alpha\times\intLeff\times$BR($\ell\ell$) for each final state.
%Adding together $N$, $N_{bck}$  and $\alpha\times\intLeff\times$BR($\ell\ell$) from each final state we use the 
%total numbers to set a limit on the cross section using a Bayesian Limit technique.
% described in~\cite{ref:bayes}. 
%We assume a flat prior (i.e. $\alpha = 1$) when obtaining the limit.
%We set an upper limit on the exclusive \Z\ cross section of $\crossexcl < 0.96$~pb at 95\%\ C.L.
We also set an upper limit on the differential cross section with respect to \yZ\ at \yZ=0 ($\dsdy$) 
using the theoretical prediction of the \yZ\ distribution~\cite{ref:motyka}.
%i.e. by assuming the theoretical description of the \Z\ production and decay kinematics to be correct. 
%While the prediction for the absolute cross section
%in Ref.~\cite{ref:motyka} and~\cite{ref:goncalves} disagree by a factor of %4.3, the ratio of $\dsdy$ to $\crossexcl$ agrees well. 
We take 0.257 as the ratio of $\dsdy$ to $\crossexcl$ and find  $\dsdy < 0.25$~pb at 95\%\ confidence level.

At hadron colliders the lepton kinematics in \ggll\ events determine the momenta of the forward (anti)protons 
through the relation $\xi(p_{1(2)}) = \frac{1}{\sqrt{s}}\sum_{i=1,2} p_T^{\ell_i}e^{+(-)\eta^{\ell_i}}$~\cite{ref:kine,ref:xi}, where
 $\xi(p_{1(2)})$ is the fractional momentum loss of the forward (backward) hadron.
In principle this relation could be used to calibrate both the
momentum scale and resolution of forward proton spectrometers.
%~\cite{fp420}.
%This is possible because the fractional momentum loss of the two protons $\xi(p_{1,2})$ can be measured both 
%directly in the spectrometers and using the lepton kinematics since
%$\xi(p_{1,2}) = \frac{1}{\sqrt{s}}\sum_{i=1,2} p_T^{l_i}e^{-(+)\eta(l_i)}$~\cite{ref:kine}.
In our eight candidate events, only one - that with $M_{\mu\mu}$ = 66.0 GeV/c$^2$ - was from a period when the Roman pot spectrometer was operational and with $\xi(\bar{p})$ in its acceptance; a track is 
observed, as expected for exclusive dilepton production.
This is an encouraging sign that if enough exclusive dilepton events are observed at the large hadron collider (LHC), they may be used to calibrate forward proton spectrometers~\cite{fp420}.
% and suggesting that this technique should be possible at the LHC. 
%(If there were any unobserved particles $\xi(\bar{p})$ from the RP track would be
%greater than that derived from the leptons, which is not the case.) 
%This is an encouraging sign that calibration of both the
%momentum scale and resolution of forward proton spectrometers at the large %hadron colider~\cite{fp420} is possible using exclusive dileptons. 

In conclusion, we have observed exclusive 
production of high mass ($\Mll > 40$ GeV/c$^2$) $e^+e^-$ and $\mu^+\mu^-$ pairs and measured a cross section that agrees with QED expectations. 
We observe no candidates for exclusive \Z\ production and put an upper limit on the photoproduction of $Z$ at a level $\approx 3,000$ times higher than
SM predictions.
It should be noted that an observation at the LHC, where the SM cross section is predicted to be 13 fb~\cite{ref:motyka}, is more promising.

We thank the Fermilab staff and the technical staffs of the participating institutions for their vital contributions. This work was supported by the U.S. Department of Energy and National Science Foundation; the Italian Istituto Nazionale di Fisica Nucleare; the Ministry of Education, Culture, Sports, Science and Technology of Japan; the Natural Sciences and Engineering Research Council of Canada; the National Science Council of the Republic of China; the Swiss National Science Foundation; the A.P. Sloan Foundation; the Bundesministerium f\"ur Bildung und Forschung, Germany; the Korean Science and Engineering Foundation and the Korean Research Foundation; the Science and Technology Facilities Council and the Royal Society, UK; the Institut National de Physique Nucleaire et Physique des Particules/CNRS; the Russian Foundation for Basic Research; the Ministerio de Ciencia e Innovaci\'{o}n, and Programa Consolider-Ingenio 2010, Spain; the Slovak R\&D Agency; and the Academy of Finland. 
%%%%%%%%% end \input{prlz_text.tex}


\begin{thebibliography}{99}
\bibitem{ref:kine} A cylindrical coordinate system is used with the $z$-axis along the 
proton beam direction; $\theta$ is the polar angle and $\phi$ is the azimuthal angle.
The transverse momentum \pt\ is the momentum perpendicular to the $z$-axis.
We define rapidity as $y=\frac{1}{2}{\rm ln}(\frac{E+p_{z}}{E-p_{z}})$ where $E$ and $p_{z}$ are the energy and momentum parallel to the $z$-axis, pseudorapidity as $\eta= -\ln \tan(\theta/2)$
and transverse energy as $E_T = E \sin\theta$.
\bibitem{ref:xi} A forward $\bar{p}$ has fractional momentum loss $\xi(\bar{p}) = 1 - \frac{p_z}{p_{beam}}$,
where $p_{beam}$ is 980~\GeVc\ and $p_{z}$ is the  $\bar{p}$ momentum after an elastic interaction.
\end{thebibliography}
\end{document}